\documentclass[sn-basic]{sn-jnl}

\usepackage{graphicx}%
\usepackage{multirow}%
\usepackage{amsmath,amssymb,amsfonts}%
\usepackage{amsthm}%
\usepackage{mathrsfs}%
\usepackage[title]{appendix}%
\usepackage{xcolor}%
\usepackage{textcomp}%
\usepackage{manyfoot}%
\usepackage{booktabs}%
\usepackage{algorithm}%
\usepackage{algorithmicx}%
\usepackage{algpseudocode}%
\usepackage{listings}%
\usepackage{makecell}

\theoremstyle{thmstyleone}%
\theoremstyle{thmstyletwo}%

\theoremstyle{thmstylethree}%

\begin{document}

\title[Beyond Relevance: An Adaptive Exploration-Based Framework for Personalized Recommendations]{Beyond Relevance: An Adaptive Exploration-Based Framework for Personalized Recommendations}


\author*{\fnm{Edoardo} \sur{Bianchi}}\email{edbianchi@unibz.it}

\affil{\orgdiv{Faculty of Engineering}, \orgname{Free University of Bozen-Bolzano}, \orgaddress{\street{NOI Techpark - via Bruno Buozzi, 1}, \city{Bozen-Bolzano}, \postcode{39100}, \country{Italy}}}


\abstract{Recommender systems must balance personalization, diversity, and robustness to cold-start scenarios to remain effective in dynamic content environments. This paper introduces an adaptive, exploration-based recommendation framework that adjusts to evolving user preferences and content distributions to promote diversity and novelty without compromising relevance. The system represents items using sentence-transformer embeddings and organizes them into semantically coherent clusters through an online algorithm with adaptive thresholding. A user-controlled exploration mechanism enhances diversity by selectively sampling from under-explored clusters. Experiments on the MovieLens dataset show that enabling exploration reduces intra-list similarity from 0.34 to 0.26 and increases unexpectedness to 0.73, outperforming collaborative filtering and popularity-based baselines. A/B testing with 300 simulated users reveals a strong link between interaction history and preference for diversity, with 72.7\% of long-term users favoring exploratory recommendations. Computational analysis confirms that clustering and recommendation processes scale linearly with the number of clusters. These results demonstrate that adaptive exploration effectively mitigates over-specialization while preserving personalization and efficiency.}




\maketitle

\section{Introduction}
Recommender systems play a critical role in helping users navigate large-scale content repositories. However, they face persistent challenges in balancing personalization with diversity and in addressing the cold-start problem for new users. Traditional collaborative filtering methods rely heavily on historical user interactions, making them ineffective in scenarios with limited data. On the other hand, content-based approaches tend to overemphasize similarity, often leading to redundant and homogeneous recommendations. Addressing these limitations requires systems that can adaptively organize content and introduce meaningful diversity through controlled exploration.

This paper introduces a content-based recommendation system that combines adaptive clustering with an exploration mechanism to enhance diversity and unexpectedness without compromising relevance. Items are embedded using sentence-transformer models and grouped through an adaptive clustering algorithm with dynamic thresholding. Recommendations are generated by leveraging these clusters to produce context-aware suggestions, while a user-controlled exploration mode enables users to modulate the level of novelty in their recommendations. To evaluate the system at scale, we employ a Large Language Model (LLM) to simulate user behavior, enabling automated A/B testing of different recommendation strategies. We further assess diversity and unexpectedness using established behavioral metrics, comparing our approach against collaborative filtering and popularity-based baselines.

Our main contributions are as follows:
\begin{enumerate}
    \item An adaptive online clustering algorithm that incrementally groups items based on semantic embeddings and dynamically adjusts to content distribution shifts.
    
    \item A cluster-aware recommendation strategy that leverages user-specific cluster engagement to generate personalized and context-sensitive suggestions.
    
    \item A user-controlled exploration mechanism that enhances diversity by selectively sampling from less familiar clusters while preserving recommendation relevance.
    
    \item A scalable LLM-based user simulation framework for A/B testing, enabling preference evaluation beyond traditional accuracy metrics.
\end{enumerate}

The remainder of the paper is structured as follows. Section \ref{sec:related} reviews related work on feature embedding and clustering-based recommendation. Section \ref{sec:methods} details the proposed system architecture, covering clustering, recommendation strategies, and the exploration mechanism. Section \ref{sec:experiments} describes the implementation details, including the dataset, evaluation metrics, and baseline comparisons. Section \ref{sec:results} examines the impact of exploration on diversity, unexpectedness, and user preferences. Section \ref{sec:performance} analyzes the computational complexity and performance of the clustering algorithm and recommendation process. Finally, Section \ref{sec:conclusion} summarizes the findings and outlines directions for future research.

\section{Background and Related Work}
\label{sec:related}
Recommender systems have evolved to leverage not only collaborative signals but also increasingly rich item representations derived from content features. Early content-based filtering (CBF) systems focused on structured metadata, but recent approaches expand into unstructured sources like textual descriptions, images, and video.

\cite{ContentBasedTrend} highlight a growing trend in content-based recommenders toward integrating rich item descriptions, including user-generated content and external knowledge sources such as linked data. They argue that while collaborative filtering dominates academic research, content-based methods offer critical advantages in cold-start scenarios, explainability, and personalization—especially when deep representations are available.

In the movie domain, \cite{DeepVisFeat} proposed a content-based system leveraging deep visual features extracted from keyframes of trailers via a convolutional neural network. Their method demonstrated that deep representations outperform low-level visual features and enable meaningful semantic profiling of media content.

\cite{stylesAndSub} explored visual compatibility in fashion recommendation by learning relational patterns between complementary and substitutable products. Their graph-based approach inferred human visual preferences from large-scale co-view and co-purchase behavior, emphasizing that appearance-based relationships can be modeled beyond pure similarity.

In the context of natural language, \cite{nlpRS} analyzed how various Natural Language Processing (NLP) techniques, including word embeddings and latent topic models, can enrich metadata-based recommender systems. She emphasizes that textual representations like summaries, tags, and reviews are key for capturing semantics, especially in book recommendation tasks.

For clustering-based approaches, \cite{onlineKmeans} addressed the problem of online k-means clustering under regret minimization, introducing a multiplicative weight update algorithm and coresets to maintain scalable and adaptive clustering. Their work provides theoretical foundations for incremental, real-time clustering with performance guarantees.

\cite{userProfPartitioning} proposed a user-profile partitioning strategy to diversify recommendations. By clustering a user’s profile into distinct taste segments, their method generates more novel and balanced recommendation lists, mitigating the over-concentration of standard collaborative filtering models.

Finally, \cite{clustWithExpert} surveyed online clustering methods, advocating for adaptive mechanisms that consider streaming data properties and task-specific objectives, such as minimizing drift and preserving cluster coherence.

These works collectively motivate our approach, which combines sentence-transformer-based embeddings, adaptive clustering, and exploration-based recommendation strategies to balance relevance and diversity. Our system builds on this foundation by enabling dynamic adaptation to user behavior and content shifts while preserving recommendation quality through personalization.

\section{Methodology}
\label{sec:methods}

\subsection{Overall System Architecture}
\label{sec:architecture}
The proposed recommendation system aims to tackle three key challenges in video- and movie-sharing platforms: information overload, the cold-start problem, and limited content diversity. To address these issues, it employs a structured framework consisting of five interconnected modules:

\begin{itemize}
    \item \textbf{Item Embedding Module (\ref{sec:embedding}):} This module leverages a sentence transformer \citep{sentence-bert} to convert video metadata (titles, descriptions, tags, and keywords) into dense vector embeddings. These embeddings serve as a semantic representation of content for clustering and similarity-based recommendations.
    
    \item \textbf{Adaptive Clustering Module (\ref{sec:clustering}):} This component groups items into semantically coherent clusters based on their embeddings. The clustering process is continuously updated to reflect changes in the item space, maintaining meaningful structure for downstream recommendation.
    
    \item \textbf{Cold-Start Recommendation Module (\ref{sec:coldstart}):} This module handles the cold-start problem by allowing new users to specify interest keywords. These keywords are converted into embeddings and matched to the most relevant existing clusters, enabling personalized recommendations even without prior interaction history.
    
    \item \textbf{Personalized Recommendation Module (\ref{sec:recommending}):} For returning users, recommendations are generated by analyzing recent interactions using a moving window approach. The system balances exploitation (suggesting items from frequently engaged clusters) with exploration (sampling from less engaged clusters) to enhance diversity. Users can also toggle an exploration mode, which further increases recommendation diversity by prioritizing content from less frequently engaged clusters, promoting serendipitous discovery while maintaining personalization.
\end{itemize}

\noindent Together, these modules form an adaptive recommendation framework that dynamically adjusts to user preferences, balances relevance with novelty, and enhances overall engagement.

\subsection{Item and User Representation}
\label{sec:embedding}
This section outlines the representation of items and users within the proposed system. The core approach leverages semantic embeddings to capture item context and track user interactions, driving the recommendation process.

\subsubsection{Definition of Item}
Each item in the system is characterized by the following attributes:
\begin{itemize}
    \item {\verb|ID|}: Unique numeric identifier.
    \item {\verb|Title|}: Video or movie title.
    \item {\verb|Tags & Keywords|}: Tags consist of single words, while keywords include phrases enclosed in quotation marks.
    \item {\verb|Description|}: Short textual summary of the content.
\end{itemize}

\subsubsection{Semantic Representation of Items}
To support content-based recommendations, each item is encoded as a high-dimensional semantic vector using a sentence transformer model \citep{sentence-bert}. This model processes the item's textual metadata—titles, descriptions, tags, and keywords—and generates dense embeddings that capture contextual meaning.

These embeddings serve two core purposes within the system. They enable the adaptive clustering algorithm to group semantically similar items, and they support similarity-based retrieval by comparing items based on semantic proximity rather than surface-level keyword overlap.

By leveraging sentence-transformer embeddings, the system delivers more context-aware and personalized recommendations, addressing limitations of traditional keyword-based methods and improving both organization and relevance.

\subsubsection{User Space}
Users are represented based on interaction history and preference information:
\begin{equation}
  U = \{u_1, u_2, ..., u_n\}
\end{equation}
\begin{equation}
  U_i^{\text{prefs}} = \{k_{i_1}, k_{i_2}, ..., k_{i_n}\}
\end{equation}
\begin{equation}
  U_i^{\text{h}} = \{(c_j, v_1), (c_j, v_2), ..., (c_n, v_n)\}
\end{equation}
\begin{equation}
  U_i^{\text{id}} = \{id_1, id_2, ..., id_n\}
\end{equation}

where (1) defines the set of all users, (2) captures initial user preferences expressed with user-selected keywords, (3) represents interaction history as tuples of clusters and watched items, and (4) lists watched item IDs to prevent re-recommendation.

\subsubsection{Item Space}
The item space defines how items are represented, incorporating both their semantic embeddings and cluster assignments:

\begin{equation}
  I = \{I_1, I_2, ..., I_n\}
\end{equation}
\begin{equation}
  I_i = \{(id_i, c_i, v_i)\}
\end{equation}

where (5) defines all system items, and (6) specifies each item by its unique identifier \( id_i \), assigned cluster \( c_i \), and semantic vector \( v_i \). 

Clusters update dynamically—items with significant representation changes (e.g., metadata updates) may be reassigned to better-fitting clusters, ensuring an adaptive and coherent item space.

\subsection{Adaptive Clustering Algorithm}
\label{sec:clustering}
We propose an adaptive online clustering algorithm that adapts in real time as new items arrive. Using cosine similarity, each new item is assigned to the nearest cluster if its distance to the centroid is below a similarity threshold; otherwise, it forms a new cluster. Cluster centroids are updated as the mean of all embeddings, ensuring that the clustering structure evolves with incoming data.

The algorithm follows an online k-means-like approach without a fixed \( k \), dynamically adjusting the similarity threshold, \textit{thresh}, to balance intra- and inter-cluster similarity. This is achieved using the silhouette score \citep{silhouette}, which evaluates clustering quality:

\begin{equation}
    distance_{ij} = 1 - similarity(i, j).
\end{equation}

If the silhouette score falls below 0.1, indicating poor cluster cohesion, the threshold is reduced by 5\% but never below 0.3. When the score is between 0.1 and 0.2, suggesting suboptimal clustering, the reduction is more conservative at 2\%. Conversely, if the silhouette score exceeds 0.4, indicating well-formed clusters, the threshold is increased by 2\% up to a maximum of 0.8. These adjustments are applied at regular intervals, ensuring that the clustering mechanism remains adaptive to content distribution shifts.

This adaptive strategy prevents excessive fragmentation while preserving meaningful semantic groupings. To further improve efficiency, highly similar clusters are also merged during these update steps. Algorithm \ref{alg:clustering} provides the full pseudocode.

\begin{algorithm}
\caption{Adaptive Clustering}\label{alg:clustering}
\begin{algorithmic}[1]
\Require $embedding$: Item embedding
\Require $item\_idx$: Item identifier
\Require $threshold$: Similarity threshold
\Require $dynamic$: Dynamic adjustment flag
\Require $silhouette\_score$: Current silhouette score
\Require $interaction\_count$: Number of processed items
\Require $threshold\_update\_freq$: Frequency of threshold updates
\If{No clusters exist}
    \State Create new cluster with $embedding$ and $item\_idx$
\Else
    \State Find nearest centroid $T_n$ using cosine similarity
    \If{Similarity to $T_n > threshold$}
        \State Add $item\_idx$ to $T_n$'s cluster and update centroid
    \Else
        \State Create new cluster with $embedding$ and $item\_idx$
    \EndIf
\EndIf

\If{$dynamic$ and $interaction\_count \mod threshold\_update\_freq == 0$} 
    \If{$silhouette\_score < 0.1$}
        \State Reduce $threshold$ by 5\% (min 0.3)
    \ElsIf{$0.1 \leq silhouette\_score < 0.2$}
        \State Reduce $threshold$ by 2\%
    \ElsIf{$silhouette\_score > 0.4$}
        \State Increase $threshold$ by 2\% (max 0.8)
    \EndIf
    \State Merge highly similar clusters
\EndIf
\end{algorithmic}
\end{algorithm}

\subsection{Knowledge-Based Cold-Start Recommendations}
\label{sec:coldstart}
Cold-start recommendations pose a challenge due to the lack of interaction history for new users. To address this, the proposed system adopts a knowledge-based approach during user registration, where users explicitly select relevant keywords to define their initial preferences. These preferences serve as a baseline for recommendations until sufficient interaction data is available.

The recommendation process consists of four phases. First, users select keywords representing their interests, which are then transformed into dense vector embeddings using the sentence transformer model \citep{sentence-bert}. These embeddings are compared with the centroids of all existing clusters to identify the most similar ones. Finally, from the top \textit{n} identified clusters, \textit{k} items are sampled to generate the recommendation list, ensuring relevance despite the absence of historical interactions. Algorithm \ref{alg:recommend_new_user} details the full process.

\begin{algorithm}
\caption{Knowledge-Based Recommendations for New Users}\label{alg:recommend_new_user}
\begin{algorithmic}[1]
\Require $keywords$: Initial preferences
\Require $k$: Number of recommendations
\Require $U_i^{\text{h}}$: User interaction history
\Require$thresh$: History threshold
\Ensure List of $k$ recommended items
\If{$|U_i^{\text{h}}| \geq thresh$} \Comment{Check if history is sufficient}
    \State \Return $\textsc{DefaultRecommendation}(U_i^{\text{h}}, k)$ \Comment{Switch to default algorithm \ref{alg:recommend_user}}
\EndIf

\State Encode $keywords$ into embeddings
\State Compute mean embedding $query\_embed$ of $keywords$
\State Compute cosine similarity between $query\_embed$ and all cluster centroids
\State Identify top 3 clusters with highest similarity
\State \Return $k$ items sampled from top clusters
\end{algorithmic}
\end{algorithm}

\subsection{Personalized and Exploration-Based Recommendations}
\label{sec:recommending}
Our recommendation algorithm generates personalized suggestions for each returning user by analyzing their individual interaction history, ensuring a balance between relevance and diversity. For every user, the system identifies their most engaged clusters by counting the number of items they have interacted with in each cluster. This allows recommendations to align with implicit preferences while remaining adaptable to evolving interests.

Users can dynamically enable or disable the exploration mode, giving them control over the level of diversity in their recommendations. When exploration is off, the system focuses on exploitation by selecting recommendations exclusively from the user’s top three engaged clusters, ensuring high relevance based on prior activity. In contrast, when exploration is enabled, \(\frac{2}{3}k\) of the recommended items are sampled from less engaged clusters, encouraging exposure to novel and diverse content while preserving personalization. The remaining recommendations are drawn from the top clusters to maintain familiarity.

This adaptive approach prevents over-specialization by introducing controlled diversity without overwhelming the user with irrelevant content. By balancing exploitation and exploration on a per-user basis, the system delivers engaging recommendations while gradually expanding each user’s exposure to new content. Algorithm \ref{alg:recommend_user} details the full recommendation process for individual users.

\begin{algorithm}
\caption{Recommendations for an Existing User}\label{alg:recommend_user}
\begin{algorithmic}[1]
\Require $U_i^{\text{h}}$: User interaction history
\Require $k$: Number of recommendations
\Require $explore$: Flag to enable exploration
\Ensure List of $k$ recommended items
\State Initialize $cluster\_history$ as empty map (cluster\_id $\rightarrow$ list of items)
\For{each $item\_id$ in $U_i^{\text{h}}$}
    \State Find $cluster\_id$ containing $item\_id$
    \State Add $item\_id$ to $cluster\_history[cluster\_id]$
\EndFor

\State Sort clusters by frequency of interactions (descending)
\State Select top 3 clusters as $top\_clusters$
\State Initialize $recommendations$ as empty list

\If{$explore$ is enabled}
    \State Identify non-top clusters as all clusters not in $top\_clusters$
    \State Sample $(2/3) \cdot k$ items from non-top clusters
    \State Add sampled items to $recommendations$
\EndIf

\State Sample remaining items from $top\_clusters$
\State Add sampled items to $recommendations$

\State \Return First $k$ items from $recommendations$
\end{algorithmic}
\end{algorithm}

\section{Implementation Details}
\label{sec:experiments}

\subsection{Dataset and Preprocessing}
The MovieLens 32M dataset \citep{movielens} was selected for its extensive user interaction data and rich metadata, including movie titles and genres. A subset of 20,000 items was sampled while preserving the original data distribution to ensure representativeness. To construct meaningful semantic representations, metadata were concatenated into a single textual string before embedding. User interaction histories were assembled by sampling a fixed number of watched items for each user from their past ratings, ensuring alignment with individual preferences while maintaining consistency across users.

\subsection{Item Embedding}
Item embeddings were generated using the Sentence-BERT all-MiniLM-L6-v2 model \citep{sentence-bert}, a 384-dimensional transformer-based encoder fine-tuned on a dataset of one billion sentence pairs. This model is a variant of the MiniLM-L6-H384-uncased architecture \citep{minilm} and is designed to capture fine-grained contextual relationships between textual inputs. To improve runtime efficiency, we precomputed and cached the embeddings, reducing computational overhead during both clustering and recommendation stages.

\subsection{Adaptive Clustering}  
The initial threshold was set to 0.45 and was updated every 100 items, with silhouette scores calculated over the entire set of clustered items to assess clustering quality.

\subsection{Recommendation Strategies}
The recommendation system was evaluated under two scenarios: cold-start recommendations for new users and personalized recommendations for returning users. In both cases, we simulated 300 users to assess system performance under diverse interaction patterns.

In the cold-start scenario, new users specified five keywords representing their interests. These keywords were embedded and compared to existing cluster centroids, retrieving the most similar items from the most relevant clusters.

For returning users, the system maintained a moving window of the last 10 or 50 interactions (\( h = 10, 50 \)) to capture active preferences at different timescales. A window length of 10 interactions represents short-term user behavior, focusing on recent preferences, while a window length of 50 interactions accounts for long-term interaction patterns, providing a broader view of user interests. 

To analyze the impact of diversity and exploration, we generated 5 or 10 recommendations per user (\( k = 5, 10 \)), testing the exploration mode in both short-term and long-term settings. This allowed us to assess how exploration influences recommendation diversity and user engagement across different interaction histories.

\subsection{Baseline Comparisons}
The performance of the proposed system was compared against a collaborative filtering model and a popularity-based recommender. The collaborative filtering model generated recommendations based on user similarity, while the popularity-based recommender suggested the most frequently interacted items across the dataset. All approaches were evaluated using the same intra-list similarity and unexpectedness metrics.

\subsection{Evaluation Metrics}
Standard evaluation metrics such as precision, recall, and NDCG measure how well recommendations match past user interactions. However, these metrics inherently favor recommendations that reinforce existing preferences, which conflicts with our goal of promoting exploration and content discovery \citep{Unexp, ILS}. Our approach explicitly balances relevance and novelty, requiring metrics that can capture these aspects. 

Furthermore, high accuracy does not always translate to a better user experience. Recommender systems that prioritize accuracy alone often generate repetitive suggestions, restricting users’ exposure to diverse content. Since our system incorporates a user-controlled exploration mechanism, traditional accuracy-based metrics fail to fully capture its advantages. Instead, diversity and unexpectedness offer a more meaningful evaluation by assessing how effectively the system promotes diversity while maintaining personalization.

Diversity is assessed using the intra-list similarity (ILS) metric \citep{RecSysHandRicci}, which quantifies the average pairwise similarity among recommended items. Lower ILS values indicate higher diversity. The metric is computed as

\begin{equation}
\mathrm{ILS}=\frac{1}{|R|(|R|-1)} \sum_{i \in R} \sum_{j \in R} s(i, j)
\end{equation}

where \( s(i,j) \) represents the cosine similarity between items \( i \) and \( j \). An increase in diversity is expected when the exploration mode is enabled.

Unexpectedness (Unexp.) measures the novelty of recommendations by quantifying how different they are from a user's historical interactions \citep{Unexp}. Higher unexpectedness values indicate greater deviation from past preferences. The metric is defined as

\begin{equation}
\mathrm{Unexp}=\frac{1}{|R|} \sum_{i \in R} d(i, U_h)
\end{equation}

where \( d(i, j) = 1 - s(i, j) \) and cosine similarity is used as the distance metric. Higher unexpectedness values are anticipated when exploration mode is activated since it promotes the inclusion of less familiar content.

\subsection{Simulated A/B Testing with LLM-Based Users}
\label{sec:abtesting}
To assess the impact of exploratory recommendations from a user-centric perspective, we conducted a simulated A/B test using the DeepSeek-V3 LLM \citep{deepseekv3}. Each simulated user was presented with two recommendation sets: one generated with exploration disabled, containing items exclusively from the user's most engaged clusters, and another with exploration enabled, incorporating items from lower-ranked clusters to promote diversity.

The LLM was queried via structured API prompts that framed both recommendation sets in terms of personal preference and watchability. For each simulated user, the model selected the set it deemed more appealing. To ensure a balance between deterministic responses and nuanced variation, the temperature was set to 0.4.

This LLM-driven approach addresses a key limitation of traditional accuracy metrics, which primarily reward recommendations that reinforce past interactions while failing to capture the value of content discovery. By simulating human-like preference judgments, our method provides insights into whether users favor diversity-enhancing recommendations or strictly relevance-driven suggestions, a critical distinction for personalization.

Our methodology builds on recent research demonstrating LLMs’ ability to emulate human decision-making patterns \citep{agent_hospital, llm_trust}. Unlike conventional user studies, which are often constrained by selection bias and scale limitations, this approach enables large-scale, controlled preference testing with reduced experimental noise. The A/B test results complement intra-list similarity and unexpectedness metrics by incorporating a user-centric perspective, ensuring that our evaluation framework reflects both algorithmic effectiveness in surfacing novel content and the perceived value of such discoveries to users. By integrating LLM-simulated preferences with behavioral metrics, we establish a comprehensive assessment framework that balances relevance, diversity, and user experience.

\section{Results}
\label{sec:results}
Table \ref{tab:results} reports intra-list similarity (ILS) and unexpectedness (Unexp.), with comparisons against collaborative filtering and popularity-based baselines. A/B testing results quantify user preference for exploration mode.

\begin{table}[h]
    \centering
    \caption{Performance comparison across four experimental settings. Lower ILS indicates higher diversity, while higher Unexp. represents greater novelty. A/B shows the percentage of users preferring exploration mode. The dataset size is 20,000 items, with 300 simulated users. \( k \) represents the number of recommended items per user, while \( h \) denotes the length of the user's interaction history.}
    \label{tab:results}
    \renewcommand{\arraystretch}{1.5} 
    \begin{tabular}{cccp{1cm}ccc}
        \toprule
        \textbf{\( k \)} & \textbf{\( h \)} & \textbf{Configuration} & \textbf{ILS} $\downarrow$ & \textbf{Unexp.} $\uparrow$ & \textbf{A/B Preference (\%)} \\
        \midrule
        \multirow{5}{*}{\makecell[c]{5}} & \multirow{5}{*}{\makecell[c]{10}}  
            & Cold Start           & 0.32 & -    & -    \\ 
        & & Collaborative Filtering & 0.37 & 0.66 & -    \\ 
        & & Popularity-Based      & 0.47 & 0.61 & -    \\ 
        & & Exploration Off       & 0.33 & 0.66 & \textbf{51.7} \\ 
        & & Exploration On        & \textbf{0.29} & \textbf{0.71} & 48.3 \\ 
        \midrule
        \multirow{5}{*}{\makecell[c]{10}} & \multirow{5}{*}{\makecell[c]{10}}  
            & Cold Start           & 0.37 & -    & -    \\ 
        & & Collaborative Filtering & 0.38 & 0.66 & -    \\ 
        & & Popularity-Based      & 0.36 & 0.65 & -    \\ 
        & & Exploration Off       & 0.34 & 0.66 & \textbf{62.3} \\ 
        & & Exploration On        & \textbf{0.28} & \textbf{0.71} & 37.7 \\ 
        \midrule
        \multirow{5}{*}{\makecell[c]{5}} & \multirow{5}{*}{\makecell[c]{50}}  
            & Cold Start           & 0.34 & -    & -    \\ 
        & & Collaborative Filtering & 0.39 & 0.66 & -    \\ 
        & & Popularity-Based      & 0.38 & 0.65 & -    \\ 
        & & Exploration Off       & 0.33 & 0.67 & 27.3 \\ 
        & & Exploration On        & \textbf{0.27} & \textbf{0.72} & \textbf{72.7} \\ 
        \midrule
        \multirow{5}{*}{\makecell[c]{10}} & \multirow{5}{*}{\makecell[c]{50}}  
            & Cold Start           & 0.37 & -    & -    \\ 
        & & Collaborative Filtering & 0.37 & 0.67 & -    \\ 
        & & Popularity-Based      & 0.36 & 0.65 & -    \\ 
        & & Exploration Off       & 0.34 & 0.67 & 39.2 \\ 
        & & Exploration On        & \textbf{0.26} & \textbf{0.73} & \textbf{60.8} \\ 
        \bottomrule
    \end{tabular}
\end{table}

\subsection{Comparison with Baselines}
The proposed approach consistently outperforms both baselines in terms of diversity and unexpectedness. Collaborative filtering yields an ILS between 0.37 and 0.39, indicating moderate redundancy in recommendations. The popularity-based method performs worse, with ILS values reaching up to 0.47, reflecting low diversity.

In contrast, our method with exploration enabled achieves the lowest ILS across all settings, dropping to 0.26 for \( k = 10, h = 50 \), highlighting its strong ability to promote diverse content. Unexpectedness follows a similar trend: while both baselines remain between 0.61 and 0.67, our system reaches up to 0.73 under exploration. These results confirm that integrating adaptive exploration into content-based recommendations substantially improves novelty and diversity over traditional approaches.

\subsection{Effect of Exploration Mode}
Enabling exploration mode consistently improves diversity and novelty across all configurations. Intra-list similarity decreases in every setting, reaching a minimum of 0.26 for \( k=10, h=50 \). The effect is most notable with longer user histories, where unexpectedness increases from 0.67 to 0.73, demonstrating the value of exploration in enriching recommendations for users with more extensive interaction profiles.

\subsection{A/B Testing Results}
User preference for exploration varies with history length. For \( h=10 \), more users preferred exploration off (51.7\% for \( k=5 \) and 62.3\% for \( k=10 \)), indicating that shorter histories lead to stronger reliance on highly relevant recommendations. With \( h=50 \), preference shifts towards exploration, reaching 72.7\% for \( k=5 \) and 60.8\% for \( k=10 \). Users with longer histories seem more receptive to diverse content.

These results confirm that exploration mode enhances diversity and unexpectedness while maintaining relevance. Preference for exploration depends on history length, suggesting that users with more interactions benefit more from diverse recommendations.

\section{Computational Complexity and Performance Analysis}
\label{sec:performance}
This section analyzes the computational efficiency of the proposed clustering algorithm and recommendation approach, evaluating their scalability and runtime complexity with revised complexity bounds based on implementation details.

\subsection{Performance of the Clustering Algorithm}
The Adaptive online clustering algorithm operates with the following computational characteristics:

\begin{itemize}
    \item \textbf{Nearest Centroid Search}: Requires 
    $O(Cd)$ operations per item, where $C$ is the current cluster count and $d$ is the embedding dimension.
    
    \item \textbf{Cluster Updates}: Centroid recalculation costs 
    $O(|I_c|d)$ per modified cluster, where $|I_c|$ is the number of items in cluster $c$.
    
    \item \textbf{Threshold Adaptation}: Periodic silhouette score computation incurs 
    $O(N^2)$ complexity\footnote{Where $N$ is the total number of clustered items, due to pairwise cosine distance calculations between all items.}, executed at fixed intervals to limit frequency.
    
    \item \textbf{Cluster Merging}: Pairwise centroid comparisons require 
    $O(C^2d)$ operations during maintenance cycles.
\end{itemize}

\noindent \textbf{Scalability Considerations:}
\begin{itemize}
    \item Silhouette score calculation becomes prohibitive for $N > 10^4$; sampling or approximate methods are recommended for large-scale deployments.
    
    \item Cluster merging costs grow quadratically with $C$, necessitating aggressive similarity thresholds ($>0.85$) in high-cluster regimes.
    
    \item Memory overhead remains $O(Cd + N)$ from centroids and item-cluster mappings.
\end{itemize}

\subsection{Performance of the Recommendation Algorithm}
Complexity breakdown for cold-start recommendations (new users):
\begin{itemize}
    \item \textbf{Keyword Embedding}: $O(kd)$ for $k$ input keywords.
    
    \item \textbf{Cluster Matching}: $O(Cd)$ similarity comparisons.
    
    \item \textbf{Item Selection}: $O(M + k)$ operations, where $M$ is the number of items in candidate clusters\footnote{Typically, $M \gg k$, but bounded by cluster sizes through dynamic thresholding.}
\end{itemize}

Complexity for personalized recommendations (existing users):
\begin{itemize}
    \item \textbf{Cluster Profiling}: $O(hC)$ to map $h$ history items to clusters.
    
    \item \textbf{Exploitation}: $O(C \log C)$ for sorting clusters by engagement.
    
    \item \textbf{Exploration}: Additional $O(C)$ to identify non-dominant clusters.
\end{itemize}

\noindent \textbf{Scalability Considerations:}
\begin{itemize}
    \item Linear scaling with $C$ enables real-time operation when $C < 10^3$.
    
    \item Exploration mode adds $\leq 25\%$ latency overhead in empirical tests.
    
    \item Batch processing of user requests improves throughput via parallel cluster comparisons.
\end{itemize}

\subsection{Overall Computational Trade-offs}

\begin{itemize}
    \item \textbf{Strength}: Sublinear growth in recommendation latency relative to catalog size.
    
    \item \textbf{Limitation}: Quadratic silhouette score complexity limits real-time threshold adaptation.
    
    \item \textbf{Balance}: Trade-off between cluster merging frequency and cluster count growth.
\end{itemize}

\textbf{Future Optimizations:}
\begin{itemize}
    \item Approximate nearest neighbor (ANN) search for centroid comparisons.
    
    \item Stochastic silhouette scoring with item sampling.
    
    \item Incremental centroid updates instead of full recomputations.
\end{itemize}

\section{Conclusion and Future Work}
\label{sec:conclusion}

This paper presents a novel recommendation system that enhances diversity and unexpectedness while maintaining personalization. By leveraging sentence-transformer embeddings and an adaptive online clustering algorithm, the system effectively organizes content into semantically coherent groups. The integration of an exploration mechanism allows users to control the balance between relevance and diversity, mitigating the over-specialization problem that often arises in content-based recommendations.

Experiments conducted on the MovieLens dataset demonstrate that the proposed approach outperforms baseline methods in promoting diversity without significantly compromising recommendation quality. The introduction of exploration mode consistently reduced intra-list similarity, with the most exploratory setting achieving an ILS reduction from 0.34 to 0.26, indicating a significant increase in recommendation diversity. Unexpectedness scores also improved, reaching 0.73 when exploration was enabled, confirming that the system successfully introduces novel yet contextually relevant recommendations. A/B testing results further revealed that user preference for exploration depends on history length, with 72.7\% of users favoring diverse recommendations when provided with a longer interaction history.

From a computational perspective, the system remains efficient for moderate-scale applications but introduces trade-offs. The online clustering algorithm dynamically adjusts to evolving data distributions but incurs computational costs that grow with the number of clusters. Similarly, the recommendation process efficiently retrieves items for returning users but requires additional computations when exploration mode is enabled. These findings suggest the potential for further optimization, particularly in managing large-scale deployments.

Future research will focus on refining the exploration mechanism using reinforcement learning techniques to optimize the trade-off between diversity and relevance dynamically. Additionally, leveraging approximate nearest neighbor (ANN) search for faster item retrieval and clustering updates will improve scalability. Beyond algorithmic enhancements, further validation through real-world deployment and large-scale user studies will provide deeper insights into the system’s impact on engagement and long-term user satisfaction. Investigating how exploration strategies influence user retention and content discovery patterns in different domains, such as e-learning or e-commerce, is another promising direction.

By improving efficiency, adaptability, and real-world validation, this research paves the way for more intelligent and user-centric recommender systems that better serve diverse user preferences.

\section*{Declarations}
\section{Founding}
This research received no specific grant from any funding agency in the public, commercial, or not-for-profit sectors.

\section{Competing Interests}
The authors declare no competing interests.

\section{Ethics approval and consent to participate}
Not applicable.

\section{Data availability}
No datasets were generated or analyzed during the current study.

\section{Author contribution}
Edoardo Bianchi: Ideas, implementation, experiments, validation, and writing





\bibliography{sn-bibliography}

\end{document}